\documentclass{article}

\usepackage{arxiv}

\usepackage[utf8]{inputenc} 
\usepackage[T1]{fontenc}    
\usepackage{hyperref}       
\usepackage{url}            
\usepackage{booktabs}       
\usepackage{amsfonts}       
\usepackage{nicefrac}       
\usepackage{microtype}      
\usepackage{graphicx}
\usepackage{float}
\usepackage{threeparttable}
\usepackage{tabularx}
\usepackage{adjustbox}
\usepackage{subcaption}
\usepackage{natbib}
\usepackage{amsmath}
\usepackage{soul,color}
\usepackage{textcomp, gensymb}
\usepackage[dvipsnames]{xcolor}

\usepackage{tikz}
\usepackage{tikz}
\usepackage{pgfplots}
\pgfplotsset{compat=1.18} 
\usetikzlibrary{external, shapes}
\usetikzlibrary{calc, matrix, positioning, chains, automata, quotes}
\usetikzlibrary{arrows.meta, arrows, backgrounds, ext.paths.ortho, ext.positioning-plus, ext.node-families.shapes.geometric, calc}

\colorlet{water}{blue!50!cyan!30}

\tikzset{
  input/.style={coordinate},
  line/.style={
    draw, -{Latex},rounded corners=3mm,
  },
  plant/.style={
    draw, rectangle, 
    text centered,
    fill=white,
    minimum height=0.75cm, 
    minimum width=1cm,
    rounded corners,
    text width=15mm,
  },
 bio/.style={
    text centered,
    minimum height=0.9cm, 
    minimum width=1cm,
    text width=15mm,
    append after command={
    \pgfextra{
        \fill[water] (\tikzlastnode.south west) rectangle ($(\tikzlastnode.south east)!0.9!(\tikzlastnode.north east)$);
        \draw (\tikzlastnode.south west) -- (\tikzlastnode.south east);
        
        \draw (\tikzlastnode.north west) -- (\tikzlastnode.south west);

        \draw (\tikzlastnode.north east) -- (\tikzlastnode.south east);
        }
        }
    },
cl/.style={
    minimum height=1.5cm, 
    minimum width=1cm,
    append after command={
    \pgfextra{
        \coordinate (A) at ($(\tikzlastnode.north) + (-1.5cm, 0cm)$);
        \coordinate (B) at ($(\tikzlastnode.north) + (1.5cm, 0cm)$);
        \coordinate (C) at ($(\tikzlastnode.south) + (1.5cm, 0cm)$);
        \coordinate (D) at ($(\tikzlastnode.south) + (-1.5cm, 0cm)$);
        \coordinate (E) at ($(\tikzlastnode.south) + (0cm, -1cm)$);
        \draw (A) -- (D) -- (E) -- (C) -- (B);
    }
}
},
 cl-square/.style={
    text centered,
    minimum height=1.2cm, 
    minimum width=1cm,
    text width=15mm,
    append after command={
    \pgfextra{
        \fill[water] (\tikzlastnode.south west) rectangle ($(\tikzlastnode.south east)!0.9!(\tikzlastnode.north east)$);
        \draw (\tikzlastnode.south west) -- (\tikzlastnode.south east);
        
        \draw (\tikzlastnode.north west) -- (\tikzlastnode.south west);

        \draw (\tikzlastnode.north east) -- (\tikzlastnode.south east);
        }
        }
    },
}

\usepackage{xurl}
\usepackage{doi}

\title{Wastewater Treatment Plant Data for Nutrient Removal System}

\author{ \href{https://orcid.org/0000-0001-7109-7944}{\includegraphics[scale=0.06]{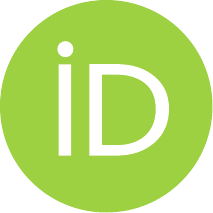}\hspace{1mm}Esmaeel Mohammadi}\thanks{Department of Chemistry and Bioscience, Aalborg University, Fredrik Bajers Vej 7H, Aalborg, 9220, North Jutland, Denmark (esmo@bio.aau.dk)}\\
    Krüger A/S\\
    Aalborg, Denmark 9210\\
    \texttt{esm@kruger.dk}\\	\texttt{esmaeel.mohammadi@veolia.com}\\
 \And
 \href{https://orcid.org/0000-0002-7856-6165}
 {\includegraphics[scale=0.06]{orcid.pdf}\hspace{1mm}
        Anju Rani}\\
	Department of Energy Technology\\
	Aalborg University\\
	Esbjerg, Denmark 6700 \\
	\texttt{aran@energy.aau.dk} \\
 \And
 \href{https://orcid.org/0000-0002-3714-5137}{\includegraphics[scale=0.06]{orcid.pdf}\hspace{1mm}Mikkel Stokholm-Bjerregaard}\\
    Krüger A/S\\
    Aalborg, Denmark 9210\\
    \texttt{mxs@kruger.dk}\\	\texttt{mikkel.stokholm-bjerregaard@veolia.com}\\
	\And
	\href{https://orcid.org/0000-0002-1297-3702}{\includegraphics[scale=0.06]{orcid.pdf}\hspace{1mm}Daniel Ortiz-Arroyo}\\
	Department of Energy Technology\\
	Aalborg University\\
	Esbjerg, Denmark 6700 \\
	\texttt{doa@energy.aau.dk} \\
	\And
	\href{https://orcid.org/0000-0003-2701-9257}{\includegraphics[scale=0.06]{orcid.pdf}\hspace{1mm}Petar Durdevic} \\
	Department of Energy Technology\\
	Aalborg University\\
	Esbjerg, Denmark 6700 \\
	\texttt{pdl@energy.aau.dk} \\
}

\renewcommand{\shorttitle}{Wastewater Treatment Plant Data for Nutrient Removal System}

\hypersetup{
pdftitle={Wastewater Treatment Plant Data for Nutrient Removal System},
pdfsubject={q-bio.NC, q-bio.QM},
pdfauthor={David S.~Hippocampus, Elias D.~Striatum},
pdfkeywords={First keyword, Second keyword, More},
}

\begin{document}
\maketitle

\begin{abstract}
This paper introduces the Agtrup (BlueKolding) dataset, collected from Denmark's Agtrup wastewater treatment plant, specifically designed to enhance phosphorus removal via chemical and biological methods. This rich dataset is assembled through a high-frequency Supervisory Control and Data Acquisition (SCADA) system data collection process, which captures a wide range of variables related to the operational dynamics of nutrient removal. It comprises time-series data featuring measurements sampled to a frequency of two minutes across various control, process, and environmental variables. The comprehensive dataset aims to foster significant advancements in wastewater management by supporting the development of sophisticated predictive models and optimizing operational strategies. By providing detailed insights into the interactions and efficiencies of chemical and biological phosphorus removal processes, the dataset serves as a vital resource for environmental researchers and engineers focused on improving the sustainability and effectiveness of wastewater treatment operations. The ultimate goal of this dataset is to facilitate the creation of digital twins and the application of machine learning techniques, such as deep reinforcement learning, to predict and enhance system performance under varying operational conditions.
\end{abstract}


\section*{Specifications Table} 
\begin{tabular}{p{4cm}p{11cm}}  
\hline
Subject                        & Time Series Forecasting and Industrial Processes Simulation\\
Specific subject area          & Deep learning methods to simulate time series data of wastewater treatment plants\\
Type of data                   & Time Series Data (CSV)\\
How data were acquired         & Data is systematically collected through the $\text{Hubgrade}^{\text{TM}}$ Performance Plant system, designed by Krüger/Veolia \cite{hubgrade}, which oversees biological and chemical treatment processes. This system captures a variety of data types, sampled every 1 to 5 minutes. Data with longer sampling intervals are up-sampled to preserve the dynamics of the process signals.\\
Data format & Raw   \\
Description of data collection & Data collection in the SCADA system for wastewater treatment is carefully designed to optimize phosphorus control operations. The logging frequency ranges between 1 and 5 minutes, with all data subsequently up-sampled to ensure the preservation of signal dynamics. The information is categorized into several types: measurements of ongoing processes, control signals generated by the phosphorus control module (excluded from predictions to allow for future system adaptability), alarm signals (or watchdogs) that prompt shifts in operational modes upon exceeding predefined thresholds, and process modes, which are represented by Boolean signals or integer codes reflecting the current operational state. The dataset is compiled into an input vector comprising 23 critical signals, including all key watchdogs, process modes, and assorted measurements. Different versions of the dataset, using both data analysis methods (principal component analysis) and process engineering knowledge, are created according to the needs for modeling.\\
Data source location & \begin{tabular}[c]{@{}l@{}}
Institution: Krüger A/S\\ 
City/Region: Søborg\\ 
Country: Denmark \\
\end{tabular}  \\
Plant location & \begin{tabular}[c]{@{}l@{}}
Institution: Bluekolding\\ 
City/Region: Kolding\\ 
Country: Denmark \\
\end{tabular}  \\
Data accessibility             & \begin{tabular}[c]{@{}l@{}}
Repository name: Mendeley Data \\ 
Data identification number: \url{} (Version 1, Version 2)\\ 
Direct URL to data: \url{} \\ 
\end{tabular} \\ 
\hline
\end{tabular}

\begin{table*}[h]
\centering
\caption{Notations for the variables used in datasets. The type describes a control variable (C), exogenous variable (E), and objective variable (O).}
\label{tab:var_notations}
\begin{tabular}{lllllc}
\hline
\textbf{Variable} & \textbf{Symbol} & \textbf{Type} & \textbf{Description} & \textbf{Unit} & \multicolumn{1}{l}{\textbf{Samples}} \\ \hline
IN\_METAL\_Q         & I  & C & Iron flow to the biology tanks         & $L/h$       & 525600 \\
T1\_O2               & O  & C & Dissolved oxygen                              & $mg/L$      & "      \\
METAL\_Q             & P  & C & polyaluminum chloride flow to the settler                & $L/h$       & "      \\
TEMPERATURE          & T  & E & Temperature of the biology tank               & $\degree C$ & "      \\
IN\_Q                & Q  & E & Flow of the wastewater to the biology tank    & $m^3/h$     & "      \\
MAX\_CF              & Cf & E & Maximum critical function percentage          & \%          & "      \\
PROCESSPHASE\_INLET  & Fi & E & Process phase at the inlet (tank 1 or 2).     & -           & "      \\
PROCESSPHASE\_OUTLET & Fo & E & Process phase at the outlet (tank 1 or 2).    & -           & "      \\
T1\_NH4              & N  & O & Ammonia concentration in biology tank 1   & $mg/L$      & "      \\
T1\_PO4              & Po & O & Phosphate concentration in biology tank 1 & $mg/L$      & "      \\ \hline
\end{tabular}
\end{table*}

\newpage
\section*{Value of the Data}
The data can be used in several applications, including but not limited to:
\begin{itemize}
\item \textbf{Process Analysis:} The time series data can be utilized to conduct detailed statistical analysis and generate visual representations of various processes within the plant. This aids in understanding and optimizing process performance.

\item \textbf{Forecasting Models:} Models such as the Autoregressive Integrated Moving Average (ARIMA) and deep learning models like Long Short-Term Memory (LSTM) networks are capable of capturing the dynamics within the time series data. These models can predict future outcomes in the nutrient removal process, offering insights into process behavior and the potential impact of integrating new technologies such as model predictive control.

\item \textbf{Simulation Environment:} Machine learning methods with recursive predictions can be employed to develop a simulation environment that mimics a digital twin of the plant. This environment is useful in correcting sensor measurement errors and training intelligent control methods, including reinforcement learning algorithms.

\item \textbf{Economic Analysis:} Analyzing time series data can also support economic studies to reduce operational costs and enhance nutrient removal and recovery. This contributes to decreasing environmental pollution from the plant’s output, providing financial and environmental benefits.
\end{itemize}

\section{Objective}
The primary objective of this study is to present the Agtrup (BlueKolding) Dataset, a comprehensive collection of time-series data from the Agtrup wastewater treatment plant (WWTP) aimed at advancing phosphorus removal techniques through biological and chemical processes. This dataset is uniquely structured to support the development of data-driven predictive models and optimization strategies that can significantly enhance the efficiency and effectiveness of wastewater treatment operations. By offering a detailed view of operational dynamics and environmental interactions, the dataset facilitates the application of machine learning and simulation technologies to create digital twins of the treatment processes. These advancements are expected to lead to improved operational control, reduced environmental impact, and enhanced economic efficiency in wastewater management \cite{mohammadi2024deep, HANSEN2022107738}.

\section{Data Description}
The dataset comprises time series data from the Agtrup wastewater treatment plant, specifically designed to study and optimize phosphorus removal. It includes different variables involved in the process from August 2021 to August 2023.

\subsection{Overview of Agtrup WWTP}
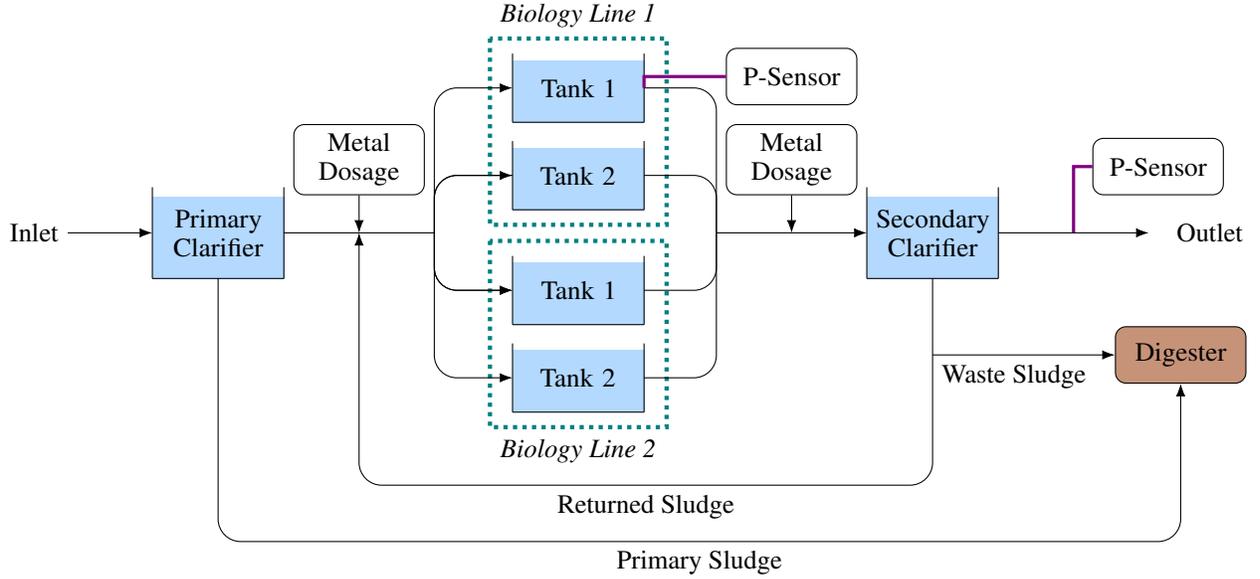
\begin{figure*}
\centering
\begin{tikzpicture}[auto, node distance=2cm,>=latex', scale=1]
    \node[input, label=left:Inlet]  (input) {};

    \node[cl-square, right of=input] (clarifier_1){Primary Clarifier};

    \node[coordinate, right =of clarifier_1] (ADL){};
    
    \coordinate (dos-1) at ($(clarifier_1.east)!.5!(ADL.west)$);
    
    \node[bio, above right of=ADL, xshift=0.5cm, yshift=-0.65cm] (A2) {Tank 2};
        
    \node[bio, above = 0.25cm of A2, label={[black, inner sep=10pt]above:\emph{Biology Line 1}}] (A1){Tank 1};

    \node[bio, below right of=ADL, xshift=0.5cm, yshift=0.65cm] (B1){Tank 1};
    \node[bio, below = 0.25cm of B1, label={[black, inner sep=10pt]below:\emph{Biology Line 2}}] (B2){Tank 2};

    \draw[teal, ultra thick, dotted] ($(A1.north west)+(-0.3,0.2)$) rectangle ($(A2.south east)+(0.3,-0.2)$);

    \draw[teal, ultra thick, dotted] ($(B1.north west)+(-0.3,0.2)$) rectangle ($(B2.south east)+(0.3,-0.2)$);

    \node[coordinate, right = 5.75cm of clarifier_1] (bio_out) {};

    \node[cl-square, right =of bio_out] (clarifier) {Secondary Clarifier};

    \coordinate (dos-2) at ($(bio_out.west)!.5!(clarifier.west)$);

    \node[plant, above = 0.5cm of dos-1] (metal1) {Metal Dosage};
    \node[plant, above = 0.5cm of dos-2] (metal2) {Metal Dosage};

    \node[right= of clarifier, label=right:Outlet] (output) {};

    \node[plant, fill=Brown!40] at ($(output.south) + (0.3cm, -1.5cm)$) (digester) {Digester};

    \coordinate (p-2) at ($(clarifier.east)!.5!(output.west)$);

    \node[plant, above = 0.25cm of metal2] (p1) {P-Sensor};
    \node[plant, above right= 0.5cm and 0.25cm of p-2] (p2) {P-Sensor};

    \draw [draw, line] (ADL) |- node {} (A1.west);
    \draw [draw, line] (ADL) |- node {} (A2.west);
    \draw [draw, line] (ADL) |- node {} (B1.west);
    \draw [draw, line] (ADL) |- node {} (B2.west);

    \draw [draw, rounded corners=3mm] (A1.east) -| node {} (bio_out);
    \draw [draw, rounded corners=3mm] (A2.east) -| node {} (bio_out);
    \draw [draw, rounded corners=3mm] (B1.east) -| node {} (bio_out);
    \draw [draw, rounded corners=3mm] (B2.east) -| node {} (bio_out);

    \draw[violet, very thick] (A1.east) |- node {} (p1);
    \draw[violet, very thick] (p-2.north) |- node {} (p2.west);

    \draw[draw, line] (clarifier.south) -- ++ (0,-2.75cm) -| (dos-1.south) node[below,pos=0.25,align=left] {Returned Sludge};

    \coordinate (intersection) at (clarifier.south |- digester.west);

    \node[below right] at (intersection) {Waste Sludge};

    \draw[draw, line] (clarifier_1.south) -- ++ (0,-3.5cm) -| (digester.south) node[below,pos=0.25,align=left] {Primary Sludge};

    \path[draw, line]
        (input) edge (clarifier_1)
        (clarifier_1) -- (ADL)
        (ADL) |- (A2)
        (bio_out) edge (clarifier)
        (clarifier) edge (output)
        (intersection) edge (digester)
        (metal1.south) edge (dos-1.north)
        (metal2.south) edge (dos-2.north)
        (ADL) |- (B1);
\end{tikzpicture}
\caption{Schematic of the phosphorus removal process in the plant with the flow lines: The iron salt is added to the inflow to the biological tanks, where P is removed, and a sensor in Tank 1 measures phosphate. A dosage of polyaluminum chloride is taken before the secondary settler to remove the remaining P, and the final phosphate concentration is measured at the outlet \cite{mohammadi2024deep}.}
\label{fig:agtrup_diagram}
\end{figure*}

The Agtrup wastewater treatment plant in Kolding, Denmark, serves 125,000 population equivalents, operating at approximately 65.5\% capacity. It employs a hybrid phosphorus removal approach, integrating chemical and biological methods. Chemical phosphorus removal (CPR) involves the addition of metal salts like iron and polyaluminum chloride at specific stages of the treatment to precipitate phosphorus into suspended solids. This occurs after the primary and before the secondary settler, respectively. Biological phosphorus removal (BPR) utilizes Phosphorus Accumulating Organisms (PAOs) that operate under alternating anaerobic and aerobic conditions. This biological process is housed between two settlers and split into two lines with two reaction tanks each, though only one tank is equipped with a phosphate sensor \cite{mohammadi2024deep, HANSEN2022107738}. A schematic of the plant's nutrient removal system is shown in Figure \ref{fig:agtrup_diagram}.

The plant faces challenges with BPR, notably the unpredictable release of phosphorus due to its complex control requirements and sensitivity to operational conditions like high nitrate concentrations, which can inhibit the necessary anaerobic processes. The plant utilizes the $\text{Hubgrade}^{\text{TM}}$ Performance Plant system developed by Krüger/Veolia to manage these processes effectively. This SCADA system controls and monitors the entire treatment process with a sampling frequency of two minutes. This system is crucial for adapting the operations to varying load conditions and ensuring effective phosphorus removal through strategic alternation between chemical and biological processes, especially during low-load periods typically occurring at night. The integrated control system helps balance the various treatment phases to optimize phosphorus removal and ensure compliance with environmental regulations. \cite{mohammadi2024deep, HANSEN2022107738}.

\subsection{Variables Description}
The dataset includes the following variables as shown in Table \ref{tab:var_notations}:
\begin{itemize}
\item IN\_METAL\_Q: Flow for the iron (Fe(III)) based chemical precipitant to the biology tanks, dosed by an on-off pump
\item T1\_O2: Dissolved oxygen levels in the biology tank 1, measured by a sensor
\item METAL\_Q\_P: Flow for the chemical precipitant (polyaluminum chloride) after the biology tanks and to the settler, dosed by an on-off pump
\item TEMPERATURE: Temperature of the biology tank, measured by a sensor
\item IN\_Q: Flow for the wastewater entering the biology tanks, measured by a flow meter
\item MAX\_CF: Maximum performance or yield (\%) of the air blowers, determined by two critical functions\footnote{These functions adjust blower output to prevent overshooting of dissolved oxygen levels, operating at maximum capacity during high nitrogen or flow demands, and reducing output when both are low. This demand-driven approach ensures efficient operation by aligning blower activity with the highest current need, enhancing treatment efficacy and energy efficiency.}: Nitrogen ($\mathrm{NH}_4$) Critical Function (NCF) and Flow Critical Function (QCF), acquired from the performance control system
\item PROCESSPHASE\_INLET: Phase code variable used in the plant to specify which tank in the biological treatment line receives the incoming wastewater flow (1 for Tank 1 and 2 for Tank 2), acquired from the performance control system
\item PROCESSPHASE\_OUTLET: Phase code variable used in the plant to indicate which tank in the biological treatment line discharges the treated wastewater (1 for Tank 1 and 2 for Tank 2), acquired from the performance control system
\item T1\_NH4: Ammonia concentration in the biology tank, measured by a sensor
\item T1\_PO4: Phosphate concentration in the biology tank, measured by a sensor
\end{itemize}

\begin{figure*}
\centering
\includegraphics[scale=1]{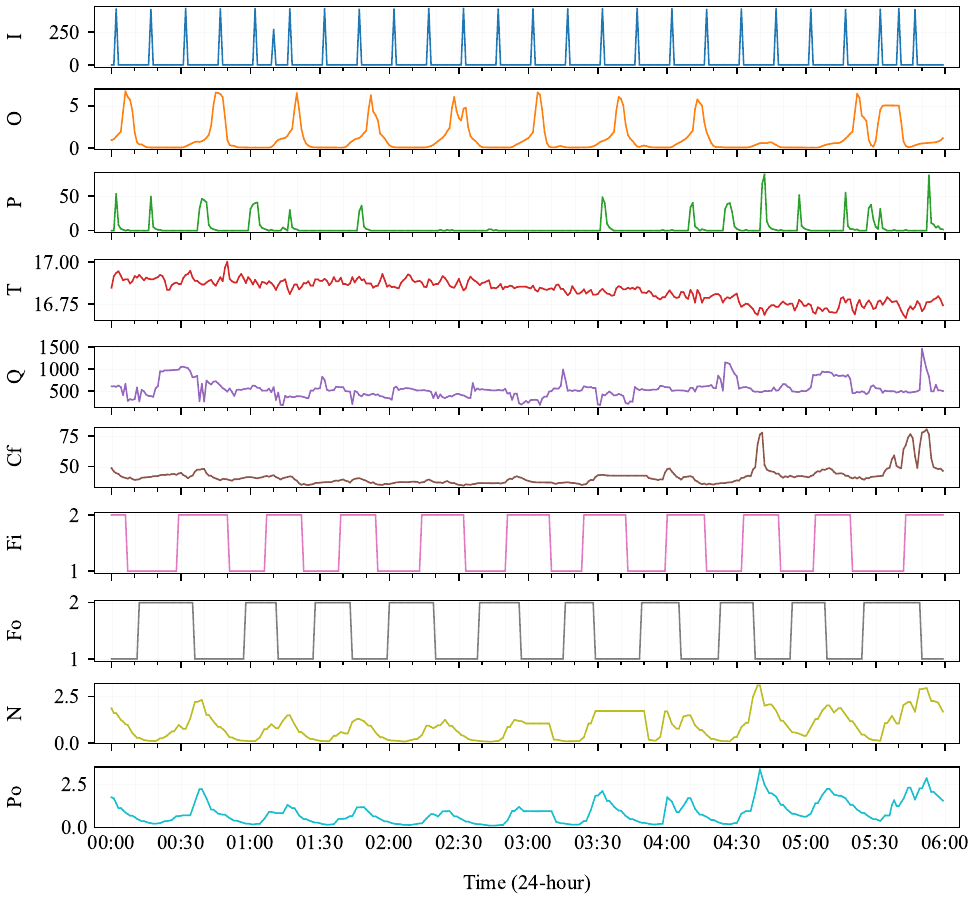}
\caption{The dynamic changes of the different variables in the biological process of wastewater treatment data}
\label{fig:24_hour}
\end{figure*}

Figure \ref{fig:24_hour} shows an example of the values and dynamics of different variables in the dataset during 6 hours.

\subsection{Variables Types}
The dataset variables impacting the nutrient removal process in the plant were categorized into three distinct groups: Control variables (C), Objective variables (O), and Exogenous Variables (E). Control variables encompass those elements of the plant that can be actively adjusted by the control system to optimize nutrient removal, such as chemical dosages and dissolved oxygen levels. Objective variables are crucial to the nutrient removal process and include measurements like phosphate and ammonia concentrations. Exogenous variables refer to factors that influence nutrient removal but remain unaffected by control or objective variable changes, including temperature, flow rates, critical functions, and phase codes. This categorization of variables can be tailored to meet the specific requirements of the study or application in question.

\subsection{Summary Statistics}
The summary statistics, shown in Table \ref{tab:summary_statistics}, provide a comprehensive view of the data characteristics for each variable included in the dataset. Below is the explanation of each column in the table:

\begin{itemize}
    \item \textbf{Variable}: The name of the data column in the dataset.
    \item \textbf{Mean}: The average data value in each variable.
    \item \textbf{Std Dev}: The standard deviation measures the amount of variation or dispersion of the data points.
    \item \textbf{Min}: The minimum value found in each variable.
    \item \textbf{25\%}: The 25th percentile, also known as the first quartile, indicates that 25\% of the data is below this value.
    \item \textbf{50\%}: The 50th percentile, or median, splits the data into two halves.
    \item \textbf{75\%}: The 75th percentile, also known as the third quartile, indicates that 75\% of the data is below this value.
    \item \textbf{Max}: The maximum value found in each variable.
\end{itemize}

\begin{table}
\centering
\caption{Summary statistics of the dataset variables.}
\label{tab:summary_statistics}
\begin{tabular}{@{}lrrrrrrr@{}}
\toprule
Variable             & Mean   & Std Dev & Min   & 25\%   & 50\%   & 75\%    & Max     \\ \midrule
IN\_METAL\_Q         & 39.98  & 99.33   & 0.00  & 0.00   & 0.02   & 0.04    & 1185.19 \\
T1\_O2               & 0.53   & 0.85    & 0.00  & 0.00   & 0.18   & 0.64    & 10.00   \\
METAL\_Q             & 4.64   & 14.12   & 0.00  & 0.01   & 0.01   & 0.02    & 149.48  \\
TEMPERATURE          & 12.74  & 3.50    & 6.07  & 9.25   & 12.80  & 15.94   & 19.52   \\
IN\_Q                & 971.74 & 520.03  & 0.00  & 583.91 & 842.77 & 1274.35 & 3988.87 \\
MAX\_CF              & 72.25  & 27.15   & 25.17 & 44.03  & 77.71  & 100.00  & 100.00  \\
PROCESSPHASE\_INLET  & 1.51   & 0.50    & 1.00  & 1.00   & 2.00   & 2.00    & 2.00    \\
PROCESSPHASE\_OUTLET & 1.49   & 0.50    & 1.00  & 1.00   & 1.00   & 2.00    & 2.00    \\
T1\_PO4              & 0.68   & 0.70    & 0.00  & 0.17   & 0.45   & 0.96    & 6.00    \\ \bottomrule
\end{tabular}
\end{table}

\section{Data Preparation Methods}
\subsection{Data Collection}
Data collection in the SCADA system for wastewater treatment is carefully designed to optimize phosphorus control operations. The logging frequency ranges between 1 and 5 minutes, with all data subsequently up-sampled to ensure the preservation of signal dynamics. The information is categorized into several types: measurements of ongoing processes, control signals generated by the phosphorus control module (excluded from predictions to allow for future system adaptability), alarm signals (or watchdogs) that prompt shifts in operational modes upon exceeding predefined thresholds, and process modes, which are represented by Boolean signals or integer codes reflecting the current operational state. The dataset is compiled into an input vector comprising 23 critical signals, including all key watchdogs, process modes, and assorted measurements. Different versions of the dataset, using both data analysis methods (principal component analysis) and process engineering knowledge, are created according to the needs for modeling \cite{mohammadi2024deep, HANSEN2022107738}.

\subsection{Data Preprocessing}
Two years of data ($D$) were collected from the SCADA system at the Agtrup plant between August 2021 and August 2023. $D$ consisted of 23 two-dimensional vectors, each representing a pair of variables at the treatment plant. Specifically, $D$ can be described as a matrix of size $n \times m$, where $n$ is the number of measurements taken over the 24 months, and $m$ is the number of variables measured. Each row of the matrix corresponds to a single measurement, and each column corresponds to a particular variable. The value of each variable is represented by the first component of the corresponding vector, while the second component represents the quality of the measurement. The quality variable takes binary values, with 0 and 1 indicating good and bad quality, respectively \cite{mohammadi2024deep, HANSEN2022107738}.

After analyzing the dataset to detect issues such as insufficient quality data, negative values, and missing values, several methods were applied for dataset treatment. The techniques used include zeroing out irrelevant data points and replacing missing and negative values with the last available positive value, a method known as Forward Filling or Last Observation Carried Forward (LOCF). This approach assumes that the last known good value is a reasonable estimate for subsequent missing or invalid values, maintaining data continuity without introducing artificial spikes or drops. Additionally, poor-quality data was substituted with the last available high-quality data point, ensuring the dataset's reliability.

\subsection{Feature Selection}
The dataset with 23 variables underwent feature selection and engineering methods after preprocessing. By replacing insufficient quality, negative, and missing data with the last available values that did not have those issues, a consistent and reliable dataset for analysis was ensured. The Pearson correlation method from the \emph{pandas} library \cite{reback2020pandas} in Python was then utilized to identify the most critical variables affecting the phosphate and ammonia concentrations in the plant. This correlation method is a statistical measure that quantifies the linear relationship or degree of association between two continuous variables \cite{FAIZI2023109}. The selected variables are described in Table \ref{tab:var_notations}.

\section*{Ethics Statement}
This research does not involve experiments, observations, or data collection related to human or animal subjects. 

\section*{Declaration of competing interest}
The authors declare that they have no known competing financial interests or personal relationships that could have appeared to influence the work reported in this paper.

\section*{Data Availability}
\href{https://data.mendeley.com/datasets/34rpmsxc4z/2} {Wastewater Treatment Plant Data for Nutrient Removal System} (Mendeley Data).

\section*{Acknowledgement}
The RecaP project has received funding from the European Union’s Horizon 2020 research and innovation programme under the Marie Skłodowska-Curie grant agreement No 956454. Disclaimer: this publication reflects only the author's view; the Research Executive Agency of the European Union is not responsible for any use that may be made of this information. We would like to thank 
Ditte Viereck Vestergaard\footnote{\texttt{divv@bluekolding.dk}} and BlueKolding A/S\footnote{\url{https://bluekolding.dk/}} for their invaluable assistance and cooperation throughout this research project and giving us consent to use and publish their data.

\bibliographystyle{unsrtnat}
\bibliography{references}  
\end{document}